\documentclass[conference]{IEEEtran}

\IEEEoverridecommandlockouts
\usepackage{cite}
\usepackage{amsmath,amssymb,amsfonts}
\usepackage{algorithm}
\usepackage[noend]{algpseudocode}
\usepackage{graphicx}
\usepackage{textcomp}
\usepackage{xcolor}
\usepackage{csquotes}
\usepackage{epstopdf}
\usepackage{tikz}
\usepackage{circuitikz}
\usepackage{graphicx,subfigure}
\usepackage{multirow}
\usepackage{url}
\usepackage{siunitx}
\usepackage{amsthm}

\usepackage{booktabs}
\usepackage{tabularx}
\usepackage{verbatim}
\usepackage{pgfplots}
\usepackage{cite}
\pgfplotsset{width=17cm,,height=4cm,compat=1.8} 
\usepackage{soul}
\usepackage[normalem]{ulem}
\usetikzlibrary{arrows}
\usepackage{verbatim}
\tikzstyle{int}=[draw, fill=blue!20, minimum height=1.5em, minimum width={width("BBBBBBBBBBB")+2pt}]
\tikzstyle{init} = [pin]
\newcommand{\argmin}[1]{\displaystyle\mathop{ \textstyle{\mbox{argmin}} }_{#1} \  }

\newcommand{\Z}{\mathbb{Z}} 
\newcommand{\R}{\mathbb{R}}

\newtheorem{theorem}{Theorem} %[section]

\newtheorem{lemma}{Lemma}

\newlength{\figwidth}
\begin{document}

\setlength{\pdfpagewidth}{8.5in}
\setlength{\pdfpageheight}{11in}
%\baselineskip 24pt
%============================================================================
%Title
%============================================================================
\title{Latency Minimization with Optimum Workload Distribution and Power Control for Fog Computing 
\vspace{-0mm}}

\author{
Saman Atapattu, Chathuranga Weeraddana,  Minhua Ding, Hazer Inaltekin and Jamie Evans
 \thanks{
S.~Atapattu and J.~Evans are with the Department of Electrical and Electronic Engineering, the University of Melbourne, Australia (e-mail: \{saman.atapattu, jse\}@unimelb.edu.au). 

C.~Weeraddana is with Department of Electronic and Telecommunication Engineering, University of Moratuwa, Moratuwa, Sri Lanka. 

M.~Ding is with Department of Electronic and Telecommunication Engineering, University of Moratuwa, Moratuwa, Sri Lanka. 

H.~Inaltekin is with School of Engineering, Macquarie University, North Ryde, NSW 2109, Australia.}
\vspace{-0.5cm}}

\maketitle

%%============================================================================
\begin{abstract}
%%============================================================================
%Smart networks containing billions of connected devices, named as Internet-of-Things (IoT), need novel computing techniques to enable efficient and reliable data processing in future wireless applications. Fog computing is one of promising paradigms to handle a large volume of time-sensitive data with minimum delay. 
This paper investigates a three-layer IoT-fog-cloud computing system to determine the optimum workload and power allocation at each layer. The objective is to minimize maximum per-layer latency (including both data processing and transmission delays) with individual power constraints. The resulting optimum resource allocation problem is a mixed-integer optimization problem with exponential complexity. Hence, the problem is first relaxed under appropriate modeling assumptions, and then an efficient iterative method is proposed to solve the relaxed but still non-convex problem. The proposed algorithm is based on an alternating optimization approach, which yields close-to-optimum results with significantly reduced complexity. Numerical results are provided to illustrate the performance of the proposed algorithm compared to the exhaustive search method. The latency gain of three-layer distributed IoT-fog-cloud computing is quantified with respect to fog-only and cloud-only computing systems. 
%utility of the proposed algorithm to infer the optimum workload allocation for the three-layer computing network under different scenarios.

%
\end{abstract}
%%============================================================================
\vspace{-2mm}
%============================================================================
%\newpage
\section{Introduction}
%============================================================================
The fifth generation ($5$G) of wireless networks and beyond are expected to support billions of connected devices, known as Internet-of-Things (IoT), by using brand-new technologies such as millimeter waves, small cells, multiple antennas, full-duplex and cooperative communications \cite{Liu2012mag,Rappaport2014ieeeproc,Atapattu13_jsac,Atapattu2019tcom}. 
%for a wide range of emerging applications. % \cite{Andrews2014jsac}.   %Billions of Internet of Things (IoT) in the 5G and beyond networks will generate enormous amount of data from a wide range of services in different domain \cite{Andrews2014jsac}.
To achieve this goal, one key challenge is the efficient processing of data generated at the network edge to meet stringent delay and reliability requirements demanded by the wide range of applications. % such as drone control, tele-health and smart traffic control. % \cite{Inaltekin18}. % ~\cite{Mukherjee2018net}. 
The cloud computing alone is often  not enough to meet all the key performance indicators in these emerging use-cases \cite{Chiang2016Fog,Gorlatova18, Inaltekin18}.  
%e.g., the latency is 1\,ms for ultra-reliable low latency communications (URLLC) applications such as autonomous vehicles, where the cloud computing alone may not be feasible~\cite{Mukherjee2018net}. 
The {\it fog computing} presents a potential solution for this problem by processing %parts of 
data locally at the fog devices. % rather than sending all raw data to the cloud. % \cite{Chiang2016Fog}.  %which distributes storage and computing resources closes to the network edge, aims to process in part workload and services locally on fog devices rather than being transmitted all data to the cloud \cite{Mukherjee2018st}.
%Fog nodes, such as routers, switches, among others can be placed close to IoT source nodes, allowing latency to be noticeably reduced compared to cloud computing. 

%The majority of differences between cloud and fog computing are attributed to the scale of hardware components associated with these computing paradigms. Cloud computing provides high availability of computing resources at relatively high power consumption, whereas fog computing provides moderate availability of computing resources at lower power consumption \cite{Jalali2016jsac}.

A typical fog computing architecture can be modeled with three layers of devices: IoT layer, fog layer, and cloud layer \cite{Gorlatova18}. %Each layer has its own data processing capability with the least computing power at the IoT layer and the most at the cloud layer ~\cite{Chiang2016Fog, Gorlatova18, Sarkar2018tcc}. The joint allocation of computing and communication resources among these three layers is of fundamental importance for minimizing data processing latency. 
To investigate different aspects of this set-up, various analytic models for control reliability~\cite{Inaltekin18}, computing delay\cite{Deng2016itj}, transmission delay\cite{Deng2016itj}, and energy consumption~\cite{Inaltekin18,Jalali2016jsac, Deng2016itj} are proposed in the existing literature and references therein.
For  device-to-device fogging\cite{Pu2016jsac}, an online task offloading problem is considered to minimize the average energy consumption. % under constraints of the user long-term incentive.
In \cite{Mao2017twc}, an online joint radio and computational resource management algorithm is developed for multi-user mobile-edge computing systems. %, where the objective is to minimize the average weighted sum power consumption. % subject to task buffer stability. % constraint.
%Moreover, to enable URLLC in mobile-edge computing, small-cell base stations (BSs) can provide  cloud-like computing and storage capabilities for end users. %' computation requests enabling ultra-low latency, where only BSs have the computation capability,  
%and therefore 
In \cite{Chen2018acm}, the resource allocation is %may be taken place 
performed between the end users and their associated small-cell base-stations when they provide cloud-like computing. % and storage functions. 
%
% \subsection{Related Work}
%
% For resource allocation in edge, fog or cloud computing, while research investigations mainly focus on two categories, power allocation \cite{Pu2016jsac,Jalali2016jsac,Mao2017twc} and delay minimization \cite{Bennis2017icc,Rodrigues2017jsac}, other  problems are also considered for bandwidth allocation \cite{Yang2018icc}, throughput maximization, energy-efficiency maximization \cite{You2017twc} and so on. 
%
% In mobile edge computing, instead of having cloud, connected small-cell base stations (BSs) provided with cloud-like computing and storage capabilities can serve end users' computation requests enabling ultra-low latency, where only BSs have the computation capability,  and therefore resource allocation may be taken place between end users and connected BSs \cite{Chen2018acm}. 
%

For a mobile-edge computing system, the user and data-set size selection is considered to minimize total energy consumption subject to computing latency in \cite{You2017twc}. 
%For a mobile-edge computation offloading system, where both mobile users and cloud have computation capabilities and the intermediate BS has no computational capability, is considered in~\cite{You2017twc}. The problem of resource allocation, in particular, the user selection and the corresponding data sizes, is formulated so that the energy consumption in the system is minimized subject to a constraint on computation latency. 
%
% For a mobile-edge computation offloading system, where both mobile users and cloud have computation capabilities and the intermediate BS has no computational capability, the optimal resource allocation, %e.g., 
% in particular, the %offloading 
% user selection and the corresponding data sizes, is formulated as a convex optimization problem in \cite{You2017twc}. This aims to minimize the energy consumption under a constraint on computation latency. 
%
A similar system consisting of single user connected to a computationally capable helper is considered in \cite{Tao2018iccws}. %Specifically, the sizes of offloaded and locally computed data are determined such that the total energy consumption for transmission and local computation is minimized under a task deadline constraint.
%
%Energy-efficient control policies are designed based on stochastic control \com{to do what?}. 
%
In \cite{Ti2017icc}, a computing system where the end-users and the cloud are connected via a BS is considered %. 
%
%model which is a combination of \cite{You2017twc,Tao2018iccws}, where the end-users and the cloud is connected via a BS, is considered. 
%The objective is 
to minimize the energy consumption. % associated to end-user computations subject to certain delay and resource constraints. %Their solution yields a policy distributing end-user computations over the cloud, designated helper and on-board processing board. 
%
%where each end-user distributes and  performs the required computation locally, remotely in the cloud, or at its designated helper. 
% In \cite{Ti2017icc}, a system model which is a combination of \cite{You2017twc,Tao2018iccws} is considered where each end-user distributes and  performs the required computation locally, remotely in the cloud (which is connected via a BS), or at its helper (which has direct link with the end user). 
%
A delay-aware and energy efficient computation offloading scheme is proposed in \cite{He2018vtc} to minimize the consumption of the non-renewable grid energy. % under delay and other network constraints. 
%
% A delay-aware energy efficient computation offloading scheme is proposed for energy harvesting enabled fog-computing enabled access points in \cite{He2018vtc}. The optimization problem is formulated to minimize the consumption of the non-renewable grid energy under delay and networks constraints. 
In \cite{Guo2018tvt}, a collaborative computation offloading is studied for cloud and mobile-edge computing to minimize
the total energy consumption. % while satisfying the maximum permissible data processing time. 
%
%Another line of research focuses on the delay minimization. 
The authors in~\cite{Bennis2017icc} considered a cloud-fog architecture to minimize the maximum computational latency.
 %, where the IoT layer does not have enough computational capability. 
%A policy to minimize the delay for an IoT-fog-cloud network is proposed in \cite{Yousefpour2017edge}. %In particular, the policy aims to minimize the service delay. 
%under maximum delay threshold.  
A trade-off between power consumption and transmission delay in a fog-cloud computing environment is investigated in \cite{Deng2016itj}. %, where the workload is split only among fog and cloud layers.
Workload allocation strategies within the fog layer subject to a power efficiency constraint are explored in \cite{Xiao2017infocom}. %Here the objective has been to minimize the response time, i.e., the round-trip transmission time and the  queuing delay.

% Note that fog computing is not intended to replace cloud computing but to compliment it, and IoT may also have power source and computational capability, in this paper, we consider a three-layer IoT-Fog-Cloud network where each layer has  computational capability. We then study optimal workload offloading between IoT, fog and cloud in order to minimize the maximum latency at each layer subject to individual maximum power constants.  For this optimization problem, to the best of our knowledge, such a more general network has not been studied in the literature.  Toward this end, we first show that this problem is non-convex.
%
We note that the fog computing is essentially to complement the cloud computing but \emph{not} to replace the cloud. In general, even the IoT layer can be equipped with certain computational capabilities. Thus, unlike in existing literature, in this paper, we consider a three-layer IoT-fog-cloud distributed computing architecture, where each layer has its \emph{own computational capacity}. We investigate the problem of splitting the workload generated by the IoT layer among the IoT, fog, and cloud. The splitting is performed so that the maximum \emph{latency} at each layer is minimized subject to individual per-layer power constraints. The resulting optimization problem is a mixed-integer program, which is intractable in general. The problem is relaxed with reasonable assumptions. An alternating optimization method is proposed with guaranteed convergence for computing a \emph{good} feasible point of the relaxed non-convex problem. For all considered empirical scenarios, negligible loss of optimality is recorded. The latency gain due to the IoT-fog-cloud computing is quantified with respect to fog-only and cloud-only systems by using the proposed method.       

% We then study optimal workload offloading between IoT, fog and cloud in order to minimize the maximum latency at each layer subject to individual maximum power constants.  

% For this optimization problem, to the best of our knowledge, such a more general network has not been studied in the literature.  Toward this end, we first show that this problem is non-convex. 
% \com{please briefly explain how we solve!}

%\vspace{-2mm}
%\newpage
%============================================================================
\section{System Model}\label{s_sys}
%============================================================================
%This section describes the  network model and the associated analytical models to characterize latencies.

\begin{figure}[t!]
\centerline{\includegraphics[width=0.5\figwidth]{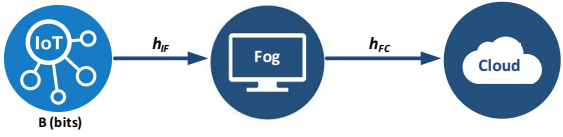}} %\vspace{-1em}
	\caption{Configuration for three-layer IoT-fog-cloud computing system.}\label{f_sys_fog}
    \vspace{-1em}
\end{figure}

We consider a network consisting of an IoT device (I), a fog node (F), and a cloud server (C), as shown in Fig.~\ref{f_sys_fog}. While the IoT device has computing resources, it may also offload some of its processing to the fog and cloud layers. We assume that the IoT device has total $B$ [bits] to be processed. It may decide to offload $m$ [bits], $m \leq B$, to the fog node and the fog node in turn may decide to offload $k$ [bits], $k \leq m$, to the cloud node. The workload distribution over the IoT, fog, and cloud layers is then $B-m$, $m-k$ and $k$ [bits], respectively. 

Processing time calculations in the proposed fog computing set-up is modelled by introducing computing powers as decision variables.
%
%An appropriate modelling approach for calculating processing times in the proposed fog computing set-up is to have computing powers as decision variables.
To this end, we start by considering a computing device with processing frequency $f$ [cycles/sec]. Assuming $E_{c}$~[joules/cycle] is required to run each computing cycle, the power consumption for processing data becomes $P_{\rm p} =  E_{c}f$ [Watts]. 
Here $E_{c}$ is considered as an intrinsic device constant depending on the underlying silicon chip technology. % to manufacture the computing device. 
Note that $P_{\rm p}$ is the total energy the computing device consumes per second. Extending this naive approach, a more refined model relating $P_{\rm p}$ and $f$ is $P_{\rm p} = a  f^\beta + b$\, [Watts],
% \begin{eqnarray}\label{e_powP}
% P_{\rm p} = a \cdot \cpu^\beta + b\,\,\,\text{[Watts]}
% \end{eqnarray}
where  $\beta$ ranges from $2.5$ to $3$, and $a$ and $b$ are  positive constants %which are 
obtained by curve fitting against empirical measurements % data %when profiling the system~
\cite{Rao2012proc, Deng2016itj}. The constant~$a$ embodies the effect of $E_{c}$ and other device parameters. Equivalently, $f = \left((P_{\rm p} - b)/a\right)^\frac{1}{\beta}$, which is the maximum data processing speed at a given computing power budget $P_{\rm p}$.    
%From $P_{\rm p}$, we get $\cpu$, the maximum processing cycles per second a computing device can execute with a specific $P_{\rm p}$, as $\cpu = \left((P_{\rm p} - b)/a\right)^\frac{1}{\beta}.$
% \begin{eqnarray}\label{e_cpu}
% \cpu = \left(\frac{P_{\rm p} - b}{a}\right)^\frac{1}{\beta}. 
% \end{eqnarray}
Now consider executing an algorithm $\mathcal{A}$ with some given complexity $\mathcal{C}(n)$, as a function of the number of input bits $n$ and in \emph{units of processor cycles} required for algorithm completion. For simplicity, assume that the complexity is linear and given by $\mathcal{C}(n) = cn$ for some positive $c$~[cycles/bit]. Thus, allocating $P_{\rm p}$ to execute $\mathcal{A}$ with $n$ bits requires a time
%processing time given by %$t_{\rm p} = c \frac{n}{\cpu} = c \,n \left(\frac{P_{\rm p} - b}{a}\right)^{-\frac{1}{\beta}}$\,[sec]. 
\begin{eqnarray}\label{e_timep}
t_{\rm p} = c \ \frac{n}{f} = c \,n \left(\frac{P_{\rm p} - b}{a}\right)^{-\frac{1}{\beta}} 
%\frac{ n}{\left(\frac{P_{\rm p} - b}{a}\right)^{\frac{1}{\beta}}}
\,\,\,\text{[sec]}. 
\end{eqnarray}
An implicit constraint here is $P_{\rm p}>b$. Let $P_{\rm tI}$, $P_{\rm tF}$ and $P_{\rm tC}$ denote the total power budgets for the IoT, fog, and cloud layers, respectively. Typically, $P_{\rm tI}\ll P_{\rm tF}\ll P_{\rm tC}$. The IoT node needs to allocate $P_{\rm tI}$ for its own data processing and communication with the fog layer. Let the local IoT power allocation for data processing and communication be denoted as $P_{\rm pI}$ and $P_{\rm cI}$, respectively, where $P_{\rm pI}+P_{\rm cI}\leq P_{\rm tI}$.  Similarly, power levels $P_{\rm pF}$ and $P_{\rm cF}$ are allocated for data processing and communication at the fog node, respectively, where $P_{\rm pF}+P_{\rm cF}\leq P_{\rm tF}$. Since the cloud only performs data processing at power $P_{\rm pC}$, we require $P_{\rm pC}\leq P_{\rm tC}$. Indexing the $a, b$ and $c$ parameters in \eqref{e_timep} with {\rm I}, {\rm F} and {\rm C}, the data processing time at each layer is
%can be given as %Denote corresponding set of parameters $\{a,b,c\}$ in \eqref{e_timep} of the IoT, fog and cloud as  $\{a_{\rm I},b_{\rm I},c_{\rm I}\}$, $\{a_{\rm F},b_{\rm F},c_{\rm F}\}$ and $\{a_{\rm C},b_{\rm C},c_{\rm C}\}$, respectively. Then the processing time at each node can be given as  

%\begin{equation}\label{e_prot}
%\begin{split}
%\text{At I}:&~t_{p,I}=c_I (B-m)\left(\frac{P_{\rm p,I} - b_I}{a_I}\right)^{-\frac{1}{\beta}}\,\,\mbox{ [sec]}\\
%\text{At F}:&~t_{p,F}=c_F\, (m-k)\left(\frac{P_{\rm p,F} - b_F}{a_F}\right)^{-\frac{1}{\beta}}\,\,\mbox{ [sec]}\\
%\text{At C}:&~t_{p,C}=c_C \,k\left(\frac{P_{\rm p,C} - b_C}{a_C}\right)^{-\frac{1}{\beta}} \,\,\mbox{ [sec]}.
%\end{split}
%\end{equation}
\begingroup
\allowdisplaybreaks
\begin{align}\label{e_prot}  \nonumber
\text{At I}:&~t_{pI}=c_I (B-m)\left(\frac{P_{\rm pI} - b_I}{a_I}\right)^{-\frac{1}{\beta}}\,\,\mbox{ [sec]},\\ \nonumber
\text{At F}:&~t_{pF}=c_F\, (m-k)\left(\frac{P_{\rm pF} - b_F}{a_F}\right)^{-\frac{1}{\beta}}\,\,\mbox{ [sec]},\\  \nonumber
\text{At C}:&~t_{pC}=c_C \,k\left(\frac{P_{\rm pC} - b_C}{a_C}\right)^{-\frac{1}{\beta}} \,\,\mbox{ [sec]}.
\end{align}
\endgroup
% \begin{equation}\label{e_prot}
% \begin{split}
% \text{At I}:&~t_{p,I}=c_I \frac{B-m}{\left(\frac{P_{\rm p,I} - b_I}{a_I}\right)^\frac{1}{\beta}};\\
% \text{At F}:&~t_{p,F}=c_F \frac{k}{\left(\frac{P_{\rm p,F} - b_F}{a_F}\right)^\frac{1}{\beta}};\\
% \text{At C}:&~t_{p,C}=c_C \frac{m-k}{\left(\frac{P_{\rm p,C} - b_C}{a_C}\right)^\frac{1}{\beta}}
% \end{split}
% \end{equation}
%
%
The IoT and fog layers is connected via a wireless link with channel gain $h_{\rm IF}$ and  bandwidth $W_{\rm IF}$. The fog communicates with the cloud via a wireless link or an optical link having channel gain $h_{\rm FC}$ and  bandwidth $W_{\rm FC}$. The throughputs for the IoT-fog and fog-cloud links are given by
\begin{equation}\label{e_rate}\nonumber 
\begin{split}
\text{I to F}:&~R_{\rm IF}=  W_{\rm IF}\log_2\left( 1 + \frac{g_{\rm IF}\, P_{\rm cI}}{N_0 W_{\rm IF}} \right) \ \mbox{[bits/sec]}, %;\,\,\,g_{\rm IF}=|h_{\rm IF}|^2\\
\\
\text{F to C}:&~R_{\rm FC}=  W_{\rm FC}\log_2\left( 1 + \frac{g_{\rm FC}\,P_{\rm cF}}{N_0 W_{\rm FC}} \right)\, \mbox{ [bits/sec]}, %;\,\,\,g_{\rm FC}=|h_{\rm FC}|^2
\end{split}
\end{equation}
where $g_{\rm IF}{=}|h_{\rm IF}|^2$, $g_{\rm FC}{=}|h_{\rm FC}|^2$, and $N_0$ is the noise spectral density. 
The communication time over each link is given by  
\begin{eqnarray} \label{e_latency}
t_{\rm c,IF} = \frac{m}{R_{\rm IF}} \,\,\mbox{ [sec]}
\,\,\,\text{and} \,\,\, 
t_{\rm c,FC} = \frac{k}{R_{\rm FC}} \,\,\mbox{ [sec]}.
\end{eqnarray}

The total latency at each stage is determined as follows. For local data processing at the IoT layer, we only have latency $T_{\rm I}$ for processing $B-m$ bits. %On the other hand, 
The latency $T_{\rm F}$ for processing $m-k$ bits at the fog layer is the sum of communication latency of $m$ bits from the IoT layer to the fog layer and the processing latency of the $m-k$ bits. %Finally, 
For the cloud, the latency $T_{\rm C}$ for processing $k$ bits is the sum of processing time at the cloud and communication latencies from the IoT layer to the fog layer and from the fog layer to the cloud layer.    
%Furthermore, if the task is routed to the cloud, that latency $T_{\rm C}$ is the sum of the communication from IoT to fog and from fog to cloud, and processing times. 
Assuming that data transmission and processing can be carried out simultaneously, % at the IoT and fog layers, 
the latencies are given by
\begin{equation}\label{e_nodet}
\begin{split}
&T_{\rm I}\left(m,P_{\rm pI}\right)=t_{\rm pI};\,\, %\\
T_{\rm F}\left(m,k,P_{\rm pF},P_{\rm cI}\right)=t_{\rm c,IF}+t_{\rm pF}\\
&T_{\rm C}\left(m,k,P_{\rm pC},P_{\rm cF}\right)=t_{\rm c,IF}+t_{\rm c,FC}+t_{\rm pC}. 
\end{split}
\end{equation}
Based on \eqref{e_nodet}, the effective system latency to complete the whole task is given by 
\begin{equation}\label{eq:T-definition}
T =\max\left(T_{\rm I},T_{\rm F},T_{\rm C} \right),
\end{equation}
where $T$ is a function of workload distribution and power allocations at IoT, fog, and cloud layers. 
%\begin{equation}\label{e_lat}
%\begin{split}
%\hspace{-1mm} T =\max\left(T_{\rm I},T_{\rm F},T_{\rm C} \right). 
%\end{split}
%\end{equation}

%\begin{equation}\label{e_lat}
%\begin{split}
%\hspace{-1mm} T\left(m,k,P_{\rm p,I},P_{\rm p,C},P_{\rm c,F},P_{\rm c,I},P_{\rm c,F}\right)=\max\left(T_{\rm I},T_{\rm F},T_{\rm C} \right). 
%\end{split}
%\end{equation}

\vspace{-0mm}
\section{Optimum Resource allocation}
\subsection{The Latency Minimization Problem}
Our goal is to discover the optimum workload distribution and power allocations at IoT, fog, and cloud layers to minimize $T$. This optimization problem can be formulated as
%\begin{align} \nonumber
%\hspace{-4mm}\mbox{minimize} & \qquad \displaystyle 
%T\left(m, k, P_{\rm pI}, P_{\rm cI}, P_{\rm pF}, P_{\rm cF}, P_{\rm pC} \right) %\, \text{in \eqref{e_lat}}
%\\ \nonumber \displaybreak[1]
%\mbox{subject to} & \qquad 0\leq m\leq B; \,\, 0\leq k\leq m \\  \label{eq:primal-1}\displaybreak[1]
%&\qquad P_{\rm pI} + P_{\rm cI} \leq P_{\rm tI} \ ; \,\, P_{\rm pF} + P_{\rm cF} \leq P_{\rm tF} \\  \nonumber \displaybreak[1] 
%&\qquad P_{\rm pI}>b_I, \ P_{\rm pF}>b_F, \ k, m\in\Z \nonumber \\
%&\qquad b_C < P_{\rm pC} \leq P_{\rm tC}, \nonumber 
%\end{align}
%---------------------------------
\begin{IEEEeqnarray}{lcl}\label{eq:primal-1}
	\mbox{minimize} & \ \ & T\left(m, k, P_{\rm pI}, P_{\rm cI}, P_{\rm pF}, P_{\rm cF}, P_{\rm pC} \right)\IEEEyessubnumber\label{eq:primal-1-1}\\
	\mbox{subject to} & \ \  & 0\leq m\leq B, \ 0\leq k\leq m  \IEEEyessubnumber\label{eq:primal-1-2}\\
	& \ \ & P_{\rm pI} + P_{\rm cI} \leq P_{\rm tI}, \ P_{\rm pF} + P_{\rm cF} \leq P_{\rm tF} \IEEEyessubnumber\label{eq:primal-1-3} \\
	& \ \ & P_{\rm pI}>b_I, \ P_{\rm pF}>b_F, \ b_C < P_{\rm pC} \leq P_{\rm tC}   \IEEEyessubnumber\label{eq:primal-1-4} \\
    & \ \ &  k, m\in\Z\IEEEyessubnumber\label{eq:primal-1-5}	\ ,
\end{IEEEeqnarray}
where $P_{\rm pI}, P_{\rm cI}, P_{\rm pF}, P_{\rm cF}, P_{\rm pC}, m$, and $k$ are the decision variables. %, and the rest are problem parameters. %$a_{\rm I}$, $b_{\rm I}$, $c_{\rm I}$, $a_{\rm F}$, $b_{\rm F}$, $c_{\rm F}$, $a_{\rm C}$, $b_{\rm C}$, $c_{\rm C}$, $\beta$, $B$, $W_{\rm IF}$, $W_{\rm FC}$, $g_{\rm IF}$, $g_{\rm FC}$, $N_0$, and $P_{\rm t,C}$ are problem parameters. 
Note that the optimization problem in \eqref{eq:primal-1} is a \emph{mixed-integer nonlinear} problem and is intractable in general~\footnote{Even in the case of mixed integer linear problems, no efficient solution methods exists, except in certain special cases, e.g., total unimodularity conditions hold, see \cite[\S~13.2]{Papadimitriou-Steiglitz-82}.}. However, a plausible strategy, especially when the solution for $m$ and $k$ are expected to be \emph{large} integers, is to relax the integer constraints~\cite[p. 307]{Papadimitriou-Steiglitz-82}. More specifically, we consider the related problem by replacing the integer constraint $k,m\in\Z$ of $\eqref{eq:primal-1}$ by $k,m\in\R$, whose epigraph problem is 
%---------------------------
\begin{align} \nonumber
\hspace{-4mm}\mbox{minimize} & \qquad t \\
\mbox{subject to} &  \qquad T_{\rm I}\left(m,P_{\rm pI}\right)\leq t, \ T_{\rm F}\left(m,k,P_{\rm pF},P_{\rm cI}\right)\leq t,  \label{eq:primal-1-relaxed} \\ \nonumber
& \qquad T_{\rm C}\left(m,k,P_{\rm pC},P_{\rm cF}\right)\leq t \\  \nonumber \displaybreak[1] 
& \qquad \mbox{Constraints \eqref{eq:primal-1-2}-\eqref{eq:primal-1-4}} \ ,
\end{align}
%---------------------------------
with decision variables $t$, $P_{\rm pI}$, $P_{\rm cI}$, $P_{\rm pF}$, $P_{\rm cF}$, $P_{\rm pC}$, $m$, and $k$ [compare with \eqref{e_nodet} and \eqref{eq:T-definition}]. 
\begin{figure*}[h!]
\small{\begin{align} \nonumber
\hspace{-4mm}\mbox{minimize} & \quad t \\ \label{eq:equ-problem}
\mbox{subject to} &  \quad  c_Ia_I^{(1/\beta)}\displaystyle\frac{B-m}{\displaystyle((1-\alpha)P_{\rm tI} - b_{\rm I})^\frac{1}{\beta}}\leq t, \ % \\  \displaybreak[1] \label{eq:equ-problem}
%&  \qquad 
\frac{m}{\displaystyle W_{\rm IF}\log_2\left( 1 + {\alpha gP_{\rm tI}} \right)}+c_Fa_F^{(1/\beta)} \frac{m-k}{\displaystyle((1-\gamma)P_{\rm tF} - b_{\rm F})^\frac{1}{\beta}} \leq t \\  \displaybreak[1]  \nonumber
&  \displaystyle\frac{m}{\displaystyle W_{\rm IF}\log_2\left( 1 + {\alpha g P_{\rm tI}} \right)} 
+\frac{k}{\displaystyle W_{\rm FC}\log_2\left( 1 + \gamma h{P_{\rm tF}} \right)}+ \frac{c_Ca_C^{(1/\beta)} \ k}{\left({P_{\rm tC} - b_{\rm C}}\right)^\frac{1}{\beta}}\leq t,
% \\ \nonumber \displaybreak[1] 
% & \qquad 
\,0\leq m\leq B, \ 0\leq k\leq m, \ %\\ \nonumber% \allowdisplaybreaks
%&\qquad 
\alpha\in[0, \alpha_{\textrm{max}}), \ \gamma\in [0, \gamma_{\textrm{max}}). 
\end{align}}
\hrule
\end{figure*}
\begin{lemma}[Total power usage]\label{re:TotPower}
At any optimal point, the power constraints of the problem \eqref{eq:primal-1-relaxed} hold with equality, i.e., $P_{\rm pI} + P_{\rm cI} = P_{\rm tI}$, $P_{\rm pF} + P_{\rm cF} = P_{\rm tF}$ and $P_{\rm pC} = P_{\rm tC}$. 
\end{lemma}
\begin{IEEEproof}
The proof is omitted due to space limitations. 
% Suppose first that the power constraint at the IoT layer holds with strict inequality, i.e., $P_{\rm p,I} + P_{\rm c,I} < P_{\rm t,I}$, for any solution of \eqref{eq:primal-1-relaxed}. Then, there exist positive scalars $\delta_1,\delta_2$ such that $P_{\rm p,I}+\delta_1 + P_{\rm c,I} +\delta_2= P_{\rm t,I}$. Let us now consider a new power split at the IoT layer given by $\bar{P}_{\rm p,I} = P_{\rm p,I} + \delta_1$ and $\bar{P}_{\rm c,I} = P_{\rm c,I} +\delta_2$. Note that the new IoT power split $\bar{P}_{\rm p,I}$ and $\bar{P}_{\rm c,I}$, together with the rest of the decision variables, is still a feasible point for problem~\eqref{eq:primal-1-relaxed}. More importantly, the new power split yields positive slack variables, denoted $s_1$, $s_2$, and $s_3$, in the first, third, and fourth inequality constraints of problem~\eqref{eq:primal-1-relaxed}, respectively. As a result, we can further decrease the optimal value $t^\star$ of the problem to yield $\bar t=t^\star-\min\{s_1,s_2,s_3\}$,  which is a  contradiction. Similar arguments can be used to show that $P_{\rm p,F} + P_{\rm c,F} = P_{\rm t,F}$ and $P_{\rm p, C} = P_{\rm t,C}$ at any solution of \eqref{eq:primal-1-relaxed}.  
\end{IEEEproof}
Using Lemma \ref{re:TotPower}, the optimization problem in \eqref{eq:primal-1-relaxed} can equivalently be reformulated as in \eqref{eq:equ-problem}, which is at the top of the next page, 
where the decision variables are $t, \alpha, \gamma, m$, and $k$. The parameters $g=g_{\rm IF}/(N_0W_{\rm IF})$ and $h=g_{\rm FC}/(N_0W_{\rm FC})$ are introduced for clarity. Although the problem~\eqref{eq:equ-problem} does not exhibit any convexity with respect to the decision variables $t, \alpha, \gamma, m$, and $k$, % As a result, the standard solvers for convex problems do not apply. However, 
the problem possesses interesting structural properties that facilitate the application of alternating optimization techniques, as we will discuss next.

% \begin{align} \nonumber
% \hspace{-4mm}\mbox{minimize} & \qquad t \\ \nonumber
% \mbox{subject to} &  \qquad d_{\rm I} \displaystyle\frac{B-m}{\displaystyle((1-\alpha)P_{\rm t,I} - b_{\rm I})^\frac{1}{\beta}}\leq t \\  \displaybreak[1] \label{eq:equ-problem}
% &  \qquad \displaystyle \frac{m}{\displaystyle W_{\rm IF}\log_2\left( 1 + {gP_{\rm t,I}}\alpha \right)}+d_{\rm F} \frac{m-k}{\displaystyle((1-\gamma)P_{\rm t,F} - b_{\rm F})^\frac{1}{\beta}} \leq t \\ \nonumber \displaybreak[1]
% & \qquad\displaystyle\frac{m}{\displaystyle W_{\rm IF}\log_2\left( 1 + {g P_{\rm t,I}\alpha} \right)} 
% +\frac{k}{\displaystyle W_{\rm FC}\log_2\left( 1 + h{P_{\rm t,F}\gamma} \right)}+d_{\rm C} {k}\leq t\\ \nonumber \displaybreak[1] 
% & \qquad 0\leq m\leq B; \qquad 0\leq k\leq m \\ \nonumber% \allowdisplaybreaks
% &\qquad \alpha\in[0,1] ; \qquad \gamma\in [0,1]\\  \nonumber%\allowdisplaybreaks
% % \qquad P_{\rm p,C} \leq P_{\rm t,C}
% &\qquad k,m\in\R \ , 
% \end{align}
% %\end{equation} 

\subsection{Solution Approach: Sequential Latency Minimization}

%Our aim is to minimize the processing time of $B$ bits of data by means of optimum allocation of resources (i.e., computing and communication power) and workload distribution over the IoT, fog and cloud layers, as formulated in \eqref{eq:equ-problem}.  
%
%Recall the question is how much the system, as a whole, can reduce the overall time to process $B$ bits if the load is shared among the IoT, the fog, and the cloud as formulated above in \eqref{eq:equ-problem}. For clarity, we use $\alpha_{\textrm{max}}=1-{b_{\rm I}}/{P_{\rm t, I}}$ and $\gamma_{\textrm{max}}=1-{b_{\rm F}}/{P_{\rm t, F}}$. Let us start by discussing briefly the key idea of the solution method. 
%
 \begin{figure*}[t!]
 {\small
 	\centering
\resizebox{.73\textwidth}{.12\textheight}{%
\begin{tikzpicture}[node distance=5cm,>=latex']
\node [int] (a) {\small$B$};
\node (b1) [above of=a,node distance=0.8cm] {IoT};
\node [int] (c) [right of=a] {\small$0$};
\node (b2) [above of=c,node distance=.8cm] {Fog};
\node [int] (d) [right of=c] {\small$0$};
\node (b3) [above of=d,node distance=.8cm] {Cloud};
\node [coordinate] (end) [right of=c, node distance=2cm]{};
\node [int] (a2) [below of=a, node distance=.75cm ] {\small$B-m^{(1)}$};
\node [int] (c2) [below of=c, node distance=.75cm ] {\small$m^{(1)}$};
\node [int] (d2) [below of=d, node distance=.75cm ] {\small$0$};
\node [int] (a3) [below of=a2, node distance=.75cm ] {\small$B-m^{(1)}$};
\node [int] (c3) [below of=c2, node distance=.75cm ] {\small$m^{(1)}-k^{(1)}$};
\node [int] (d3) [below of=d2, node distance=.75cm ] {\small$k^{(1)}$};
\node [int] (a4) [below of=a3, node distance=.75cm ] {\small$B-m^{(1)}-m^{(2)}$};
\node [int] (c4) [below of=c3, node distance=.75cm ] {\small$m^{(1)}+m^{(2)}-k^{(1)}$};
\node [int] (d4) [below of=d3, node distance=.75cm] {\small$k^{(1)}$};
\path[->] (a2) edge node[above] {\small$m^{(1)}$} (c2);
\path[->] (c3) edge node[above] {\small$k^{(1)}$} (d3);
\path[->] (a4) edge node[above] {\small$m^{(2)}$} (c4);
\node (s1) [right of=d2,node distance=3.4cm] {\small End of Stage 1};
\node (s1) [right of=d3,node distance=3.4cm] {\small End of Stage 2};
\node (s1) [right of=d4,node distance=3.4cm] {\small End of Stage 1};
%\path[->] (a) edge node {$v$} (c);
%\draw[->] (c) edge node {$p$} (end) ;
\end{tikzpicture}
}
	\vspace{-0em}
	\caption{Proposed Sequential Optimization Method: Alternating Stage $1$ and Stage $2$}
	\label{fig:Stage_evolution}\vspace{-1em}
    }
\end{figure*}
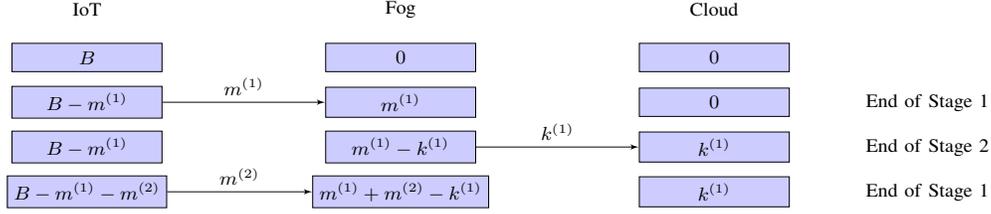
 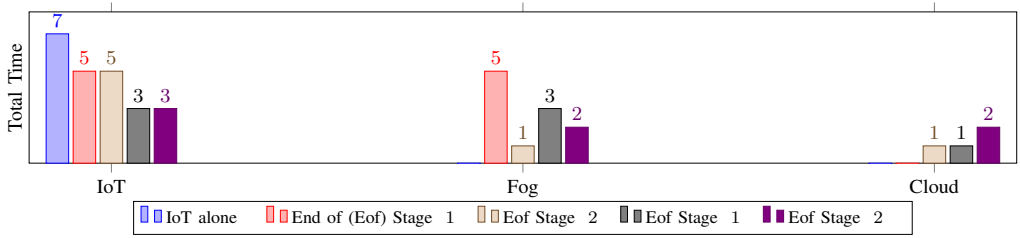
\begin{figure*}[t!]
 {\small
 	\centering
 	\resizebox{.75\textwidth}{.131\textheight}{%
\begin{tikzpicture} 
\begin{axis}[  ytick=\empty,  ybar,
    enlargelimits=0.1,
    legend style={at={(0.5,-0.25)},
     anchor=north,legend columns=-1}, 
    ylabel={Total Time},
    symbolic x coords={IoT,Fog,Cloud},
    xtick=data,
    nodes near coords,
    nodes near coords align={vertical}, ymin=0.75, ymax=7.5
    ]
\addplot coordinates {(IoT,7) (Fog,0) (Cloud,0)};
\addplot coordinates {(IoT,5) (Fog,5) (Cloud,0)};
\addplot coordinates {(IoT,5) (Fog,1) (Cloud,1)};
 \addplot coordinates {(IoT,3) (Fog,3) (Cloud,1)};
  \addplot coordinates {(IoT,3) (Fog,2) (Cloud,2)};
%    \addplot coordinates {(IoT,2.5) (Fog,2.5) (Cloud,2)};
\legend{\footnotesize{IoT alone $\quad$} , \footnotesize{End of (Eof) Stage $ \ 1 \quad $}, \footnotesize{Eof Stage $ \ 2 \quad $},\footnotesize{Eof Stage $ \ 1 \quad $}, \footnotesize{Eof Stage $ \ 2 \quad $}}
\end{axis}
\end{tikzpicture}
}
\vspace{0em}
\caption{Proposed Sequential Optimization Method: (Total Processing $+$ Communication) Time Evolution.}
\label{fig:time_evolution}\vspace{-0em}
}
\end{figure*}
%
%
%\subsection{Key Idea of the Solution Method}

For clarity, let $\alpha_{\textrm{max}}=1-{b_{\rm I}}/{P_{\rm tI}}$ and $\gamma_{\textrm{max}}=1-{b_{\rm F}}/{P_{\rm tF}}$. The key step in our method to solve \eqref{eq:equ-problem} is to decompose the latency minimization~\eqref{eq:equ-problem} into two manageable sub-problems that can be solved sequentially in two stages until a convergence criterion is satisfied. The main  idea is illustrated in Fig.~\ref{fig:Stage_evolution}. 
%
%Our approach can be split into two stages of optimizations, which are performed in an alternating manner until some convergence criterion is satisfied [\cf Fig.~\ref{fig:Stage_evolution}]. 
%
Let us consider the first iteration. At stage $1$, the IoT layer solves for the optimum number of bits $m^{(1)}$ (out of $B$) that it can assign to the fog layer, so that the overall time to process $B$ bits at the IoT and fog layers is jointly minimized. In particular, the following problem is solved at stage $1$ of iteration $1$:
%\begin{align} \nonumber
%\hspace{-4mm}\mbox{minimize} & \qquad t \\ \nonumber
%\mbox{subject to} &  \qquad \left[\displaystyle\frac{Bc_Ia_I^{(1/\beta)}}{\displaystyle((1-\alpha)P_{\rm t,I} - b_{\rm I})^\frac{1}{\beta}}\right]-\left[\displaystyle\frac{c_Ia_I^{(1/\beta)}}{\displaystyle((1-\alpha)P_{\rm t,I} - b_{\rm I})^\frac{1}{\beta}}\right] \ m \ \leq t \\  \displaybreak[1] \label{eq:equ-problem-stg-1}
%& \qquad  \left[\frac{1}{\displaystyle W_{\rm IF}\log_2\left( 1 + {gP_{\rm t,I}}\alpha \right)}+ \frac{c_Fa_F^{(1/\beta)}}{\displaystyle(P_{\rm t,F} - b_{\rm F})^\frac{1}{\beta}}\right] \ m  \ \leq t \\ \nonumber \displaybreak[1]
%& \qquad 0\leq m\leq B \\ \nonumber \displaybreak[1]
%&\qquad \alpha\in[0,1]  \ , 
%\end{align}
\begin{align} \nonumber
\hspace{-4mm}\mbox{minimize} & \quad t \\ \nonumber
\mbox{subject to} &  \quad c_Ia_I^{(1/\beta)}\left[\displaystyle\frac{B-m}{\displaystyle\big([1-\alpha]P_{\rm tI} - b_{\rm I}\big)^\frac{1}{\beta}}\right] \ \leq t \\ \nonumber \displaybreak[1] 
&  \left[\frac{m}{\displaystyle W_{\rm IF}\log_2\left( 1 + {\alpha gP_{\rm tI}}\right)}+ \frac{c_Fa_F^{(1/\beta)} \ m}{\displaystyle(P_{\rm tF} - b_{\rm F})^\frac{1}{\beta}}\right] \ \leq t \\  \displaybreak[1]\label{eq:equ-problem-stg-1}
& \quad 0\leq m\leq B, 
% \\ \nonumber \displaybreak[1]
% &\quad 
\quad 0\leq \alpha \leq \alpha_{\textrm{max}}  \ , 
\end{align}
where the decision variables are $t, m,\alpha$ only and the solution is $(t^{(1)},m^{(1)},\alpha^{(1)})$. The optimization problem \eqref{eq:equ-problem-stg-1} is simply~\eqref{eq:equ-problem} with $k=0$, $\gamma=0$, and without the $3$rd constraint.
The idea is depicted in Figs.~\ref{fig:Stage_evolution} and~\ref{fig:time_evolution}. In this example, the IoT layer requires $7$ [secs] to process $B$ bits alone. After solving stage $1$ optimization problem \eqref{eq:equ-problem-stg-1}, the IoT and fog layers together require only $5$ [secs] to process $B$ bits, which is around $30$\% latency improvement. The latency at stage $1$ includes the processing time of $\left(B{-}m^{(1)}\right)$ bits at the IoT \emph{or} communication and processing times of $m^{(1)}$ bits at the fog. %Note that the $5$s aforementioned is the processing time of $\left(B-m^{(1)}\right)$ bits at the IoT. The time is also identical to the aggregate time to communicate $m^{(1)}$ bits from the IoT to the fog and the processing of $m^{(1)}$ bits at the fog.

In stage $2$ of iteration $1$, the fog solves for the optimum number of bits $k^{(1)}$ that can be assigned to the cloud, so that  the overall time to process the already assigned $m^{(1)}$ bits at the fog and cloud are jointly minimized, as illustrated in Fig.~\ref{fig:Stage_evolution}. In other words, the following is solved at stage $2$ of iteration~$1$: %In other words, the fog checks whether a part of its local processing burden of $m^{(1)}$ bits can be assigned to the cloud to minimize the overall processing time. The related problem is simply given by
\begin{align} \nonumber
\hspace{-4mm}\mbox{minimize} & \qquad s \\ \nonumber \displaybreak[1]
\mbox{subject to} &  \qquad  \frac{m^{(1)}}{\displaystyle W_{\rm IF}\log_2\left( 1 + {gP_{\rm t,I}}\alpha^{(1)} \right)} \\ \nonumber \displaybreak[1]
&  \hspace{7.5em} +\displaystyle\frac{c_Fa_F^{(1/\beta)}\left(m^{(1)}-k\right)}{\displaystyle((1-\gamma)P_{\rm t,F} - b_{\rm F})^\frac{1}{\beta}} \ \leq s \\  \nonumber\displaybreak[1] 
& \qquad  \frac{m^{(1)}}{\displaystyle W_{\rm IF}\log_2\left( 1 + {gP_{\rm t,I}}\alpha^{(1)} \right)} + \frac{c_Ca_C^{(1/\beta)} \ k}{\displaystyle(P_{\rm t,C} - b_{\rm C})^\frac{1}{\beta}} \\ \nonumber \displaybreak[1]
&   \hspace{6.5em}+ \frac{k}{\displaystyle W_{\rm FC}\log_2\left( 1 + {hP_{\rm t,F}}\gamma \right)} %+ \frac{c_Ca_C^{(1/\beta)} \ k}{\displaystyle(P_{\rm t,C} - b_{\rm C})^\frac{1}{\beta}}\right]  
\ \leq s \\ \displaybreak[1]
& \qquad 0\leq k\leq m^{(1)}, \ % \\ \nonumber \displaybreak[1]
%&
0\leq \gamma\leq \gamma_{\textrm{max}} \ ,   \displaybreak[1] \label{eq:equ-problem-stg-2}
\end{align}
where the decision variables are $s, k,\gamma$ and the solution is $\left(s^{(1)}, k^{(1)},\gamma^{(1)}\right)$. The problem~\eqref{eq:equ-problem-stg-2} is simply the problem~\eqref{eq:equ-problem}, while leaving its first constraint out and considering $m=m^{(1)}$ and~$\alpha=\alpha^{(1)}$.
Fig.~\ref{fig:time_evolution} illustrates a situation, where the the aggregate time for communication of $m^{(1)}$ bits from the IoT layer to the fog  layer and the processing of ($m^{(1)}-k^{(1)})$ bits at the fog layer is $1$ [sec]. So is the aggregate time to communicate and process $k^{(1)}$ bits at the cloud layer. %from the fog layer to the cloud and the processing of $k^{(1)}$ bits at the cloud. 
We observe, however, that the total latency is still $5$ [secs] since the latency at the IoT layer does not improve after solving the second stage optimization problem. This is the status at the end of the first iteration. 

To further improve the latency bottleneck at the IoT layer, in the next iteration, we revert to the stage $1$ optimization problem again, which results in $3$ [secs] to process $(B-m^{(1)}-m^{(2)})$ bits at the IoT layer after solving \eqref{eq:equ-problem-stg-1} with the updated workload distribution. See Fig.~\ref{fig:time_evolution}. The aggregate communication and processing time of $(m^{(1)}+m^{(2)})$ bits at the fog layer is now equal to $3$ [secs], without any change in the cloud latency. Then, the solution method again proceeds to stage $2$ of the second iteration. The process is thus repeated. Fig.~\ref{fig:time_evolution} shows the evolution of the aggregate time to process the data at different layers for a case in which the two-stage optimization procedure iterates twice. 

%Recall at the end of stage $2$, the processing time at IoT is $5$s, where as the ag{\rm c,I}^{(i+1)}g be the communication power used by IoT in $(i+1)$th iteration, $i=1,2,\cdots$. Then $$regate time for communication of $m^{(1)}$ bits from IoT to fog and the processing of $(m^{(1)}-k^{(1)})$ bits at the fog is $1$s. This difference of the times suggests that the IoT can send some more $m^{(2)}$ bits towards the fog to further minimize the aggregate time to process the bits, \cf Fig.~\ref{fig:Stage_evolution}. In other words, the solution method revert to the stage $1$ again, which results $3$s to process $(B-m^{(1)}-m^{(2)})$ bits at the IoT, see Fig.~\ref{fig:time_evolution}. Moreover the same time is spent to communicate $(m^{(1)}+m^{(2)})$ bits from the IoT to the fog  and to process those bits at the fog, see Fig.~\ref{fig:time_evolution}. The solution method again reverts to stage~$2$, and thus the process is repeated. Fig.~\ref{fig:time_evolution} shows the evolution of aggregate time to process the bits at different nodes for a case where the two stage optimizations is repeated twice. 

%Our solution approach explained above allows us to compute a monotonically decreasing sequence of objective values for the problem \eqref{eq:equ-problem}. 
Next, we will discuss the properties of the two-stage optimization procedure.

%sequence of upper bounds (stage~$1$ optimization) and a sequence of lower bounds (stage~$2$ optimization) on the optimal value of problem~\eqref{eq:equ-problem} that are asymptotically tight. 

\subsection{Basis for the Two-Stage Optimization Procedure}\label{subsec:Basis-for-the-Two-Stage-Optimization-Procedure}

Implementation of the proposed solution technique requires solving the stage $1$ and $2$ optimization problems sequentially in an iterative manner. In other words, in any iteration, first the stage $1$ optimization is performed followed by the stage~$2$ optimization.
%    \begin{remark}[Communication power expenditure in consecative iterations of stage~1 optimizations]\label{re:compare_s_n_t_n} 
%   	Let $P_{{\rm c,I}}^{(i+1)}$ be the communication power used by IoT in $(i+1)$th iteration, stege~1, $i=1,2,\ldots$. Then $P_{\rm c,I}^{(i+1)}\geq P_{\rm c,I}^{(i)}$ for all $i$.
%   \end{remark}
%   \begin{proof}
% Let $t^{(i)}$ denote that optimal value of stege~1 optimization in the $i$th iteration. If $P_{\rm c,I}^{(i+1)}< P_{\rm c,I}^{(i)}$, we can show that, together with $P_{\rm c,I}^{(i+1)}$, we can construct a feasible point to the stage~1 optimization in the $i$th iteration, which yields an objective value which is smaller than $t^{(i)}$. This is impossible. Details are excluded. 
%  \end{proof}
%
The stage~$1$ optimization in the $i$th iteration is generally  expressed as
\begin{align} \nonumber
\hspace{-4mm}\mbox{minimize} & \quad t \\ \nonumber
\mbox{subject to} &  \quad a_{1i}(\alpha)-b_{1i}(\alpha)m \ \leq t, \ d_{1i}(\alpha) + c_{1i}(\alpha)m \ \leq t \\ \label{eq:it_i_stage_1} \displaybreak[1]
& \textstyle\quad 0\leq m\leq B-\sum_{j=0}^{i-1}m^{(j)} \\ \nonumber \displaybreak[1] 
&\textstyle\quad 0\leq \alpha\leq \alpha_{\textrm{max}}-\sum_{j=0}^{i-1}\alpha^{(j)}  \ , 
\end{align}
where the decision variables are $t, m,\alpha$ and the solution is $(t^{(i)},m^{(i)},\alpha^{(i)})$.~\footnote{The formulation ensures that the cumulative number of bits transmitted from IoT by the end of stage~1 of $i$th iteration is no smaller than that of $(i-1)$th iteration and so is for the communication power.} The problem parameters $a_{1i},b_{1i},c_{1i}$ and $d_{1i}$ for $i=1,2,\ldots$ are defined in \eqref{eq:1i_para} on the next page. %as follows:

% \com{
% {\em 
% {\bf Questions:}
% \begin{enumerate}
% \item Can we add the powers as in this formulation? In other words, do we always need to increase communication power as we send more bits to the fog layer? Seems logical but requires some proof?
% \item Given some workload allocation from fog to cloud as a result of the solution of the second stage optimization problem, do we always increase the number of bits allocated to the fog layer? Again, seems intuitive but will require a little proof. Given these two properties are correct, then at the $i$th iteration we write the problem as in \eqref{eq:it_i_stage_1} and find $\alpha$ and $m$ increments. 
% \end{enumerate}}
% } 

%@@@@@@@@@@@@@@@@@@@@@
\begin{figure*}
\small{\begin{align} \nonumber
& a_{1i}(\alpha)=\left[\displaystyle\frac{c_Ia_I^{(1/\beta)}\left(B-\sum_{j=0}^{i-1}m^{(j)}\right)}{\displaystyle\left(\left[1-\textstyle\sum_{j=0}^{i-1}\alpha^{(j)}-\alpha\right]P_{\rm t,I} - b_{\rm I}\right)^\frac{1}{\beta}}\right];\,
    c_{1i}(\alpha)=\frac{1}{\displaystyle W_{\rm IF}\log_2\left( 1 + {gP_{\rm t,I}}\left[\textstyle\sum_{j=0}^{i-1}\alpha^{(j)}+\alpha\right] \right)} 
  + \frac{c_Fa_F^{(1/\beta)} }{\displaystyle\left(\left[1-\textstyle\sum_{j=0}^{i-1}\gamma^{(j)}\right]P_{\rm t,F} - b_{\rm F}\right)^\frac{1}{\beta}};\\
 & b_{1i}(\alpha)=\frac{a_{1i}(\alpha)}{B-\sum_{j=0}^{i-1}m^{(j)}};\,
d_{1i}(\alpha)=\frac{\sum_{j=0}^{i-1}m^{(j)}}{\displaystyle W_{\rm IF}\log_2\left( 1 + {gP_{\rm t,I}}\left[\textstyle\sum_{j=0}^{i-1}\alpha^{(j)}+\alpha\right] \right)} 
+ \frac{c_Fa_F^{(1/\beta)} \ \sum_{j=0}^{i-1}\left(m^{(j)}-k^{(j)}\right)}{\displaystyle\left(\left[1-\textstyle\sum_{j=0}^{i-1}\gamma^{(j)}\right]P_{\rm t,F} - b_{\rm F}\right)^\frac{1}{\beta}}. \label{eq:1i_para}
\end{align}}
\hrule
\end{figure*}
%@@@@@@@@@@@@@@@@@@@@
% \begin{equation}\label{eq:a1i_para_b1i_para}
% \begin{split}
% 	a_{1i}(\alpha)&=\left[\displaystyle\frac{c_Ia_I^{(1/\beta)}\left(B-\sum_{j=0}^{i-1}m^{(j)}\right)}{\displaystyle\left(\left[1-\textstyle\sum_{j=0}^{i-1}\alpha^{(j)}-\alpha\right]P_{\rm t,I} - b_{\rm I}\right)^\frac{1}{\beta}}\right],\\
%     b_{1i}(\alpha)&=\frac{a_{1i}(\alpha)}{B-\sum_{j=0}^{i-1}m^{(j)}},
% \end{split}
% \end{equation}
% \begin{equation}\label{eq:d1i_para}
% \begin{split}
% d_{1i}(\alpha)=&\frac{\sum_{j=0}^{i-1}m^{(j)}}{\displaystyle W_{\rm IF}\log_2\left( 1 + {gP_{\rm t,I}}\left[\textstyle\sum_{j=0}^{i-1}\alpha^{(j)}+\alpha\right] \right)} \\
% &+ \frac{c_Fa_F^{(1/\beta)} \ \sum_{j=0}^{i-1}\left(m^{(j)}-k^{(j)}\right)}{\displaystyle\left(\left[1-\textstyle\sum_{j=0}^{i-1}\gamma^{(j)}\right]P_{\rm t,F} - b_{\rm F}\right)^\frac{1}{\beta}},
% \end{split}
% \end{equation}
% %and
% \begin{equation}\label{eq:c1i_para}
% \begin{split}
% c_{1i}(\alpha)=&\frac{1}{\displaystyle W_{\rm IF}\log_2\left( 1 + {gP_{\rm t,I}}\left[\textstyle\sum_{j=0}^{i-1}\alpha^{(j)}+\alpha\right] \right)} \\
%  & + \frac{c_Fa_F^{(1/\beta)} }{\displaystyle\left(\left[1-\textstyle\sum_{j=0}^{i-1}\gamma^{(j)}\right]P_{\rm t,F} - b_{\rm F}\right)^\frac{1}{\beta}}.
% \end{split}
% \end{equation}
%
%@@@@@@@@@@@@@@@@@@@@@@@@@@
For a \emph{fixed} $\alpha$, the solution of \eqref{eq:it_i_stage_1} can easily be computed by considering the \emph{intersection} of the lines $a_{1i}(\alpha)-b_{1i}(\alpha)m$ and $d_{1i}(\alpha) + c_{1i}(\alpha)m$. %More specifically, those points are given by $m^{(i)}(\alpha) = \frac{a_{1i}(\alpha)-d_{1i}(\alpha)}{b_{1i}(\alpha)+c_{1i}(\alpha)}$ and $t^{(i)}(\alpha) = \frac{c_{1i}(\alpha)a_{1i}(\alpha) + d_{1i}(\alpha)b_{1i}(\alpha) }{b_{1i}(\alpha)+c_{1i}(\alpha)}$. 
Specifically,  $m^{(i)}(\alpha) = \frac{a_{1i}(\alpha)-d_{1i}(\alpha)}{b_{1i}(\alpha)+c_{1i}(\alpha)}$ and $t^{(i)}(\alpha) = \frac{c_{1i}(\alpha)a_{1i}(\alpha) + d_{1i}(\alpha)b_{1i}(\alpha) }{b_{1i}(\alpha)+c_{1i}(\alpha)}$.
% \begin{equation}\label{eq:optimal_m_in_it_i_stage_1}
% m^{(i)}(\alpha) = \frac{a_{1i}(\alpha)-d_{1i}(\alpha)}{b_{1i}(\alpha)+c_{1i}(\alpha)} 
% \end{equation}
% %and 
% \begin{equation}\label{eq:optimal_t_in_it_i_stage_1}
% t^{(i)}(\alpha) = \frac{c_{1i}(\alpha)a_{1i}(\alpha) + d_{1i}(\alpha)b_{1i}(\alpha) }{b_{1i}(\alpha)+c_{1i}(\alpha)}.
% \end{equation}
Based on these, $\alpha^{(i)}$ which solves \eqref{eq:it_i_stage_1} is given by
\begin{equation}\label{eq:alpha_it_i_stage_1}
\alpha^{(i)} = \argmin{ 0 \leq   \alpha  \leq  \alpha_{\textrm{max}}-\sum_{j=0}^{i-1}\alpha^{(j)}} \ t^{(i)}(\alpha),
\end{equation}
which can be computed by using a scalar grid search over the range of $\alpha$. Substituting $\alpha^{(i)}$  yields the solutions $m^{(i)}$ and $t^{(i)}$ for \eqref{eq:it_i_stage_1}, respectively.

%Similar arguments can be used to show that the stage $2$ optimization in the $i$th iteration is
Similarly, the stage $2$ optimization in the $i$th iteration is
\begin{align}
\nonumber
\hspace{-4mm}\mbox{minimize} & \quad s \\ \nonumber
\mbox{subject to} &  \quad L_{2i} + a_{2i}(\gamma)-b_{2i}(\gamma)k \ \leq s \\  \displaybreak[1] \nonumber
& \quad L_{2i}+ d_{2i}(\gamma) + c_{2i}(\gamma)k \ \leq s \\ \label{eq:it_i_stage_2} \displaybreak[1] 
& \textstyle\quad 0\leq k\leq \sum_{j=0}^{i}m^{(j)}-\sum_{j=0}^{i-1}k^{(j)} \\ \nonumber \displaybreak[1]
&\textstyle\quad 0\leq \gamma\leq \gamma_{\textrm{max}}-\sum_{j=0}^{i-1}\gamma^{(j)}  \ , 
\end{align}
where the decision variables are $s, k,\gamma$ and the solution is $(s^{(i)},k^{(i)},\gamma^{(i)})$. The problem parameters $a_{2i},b_{2i},c_{2i}$, $d_{2i}$ and $L_{2i}$ for $i=1,2,\ldots$ are given in \eqref{eq:2i_para} shown on the next page. 
%@@@@@@@@@@@@@@@@@@@@@
\begin{figure*}
\small{\begin{align} \nonumber
& a_{2i}(\gamma)=c_Fa_F^{(1/\beta)}\left[\displaystyle\frac{\left(\sum_{j=0}^{i}m^{(j)}{-}\sum_{j=0}^{i-1}k^{(j)}\right)}{\displaystyle\left(\left[1{-}\textstyle\sum_{j=0}^{i-1}\gamma^{(j)}{-}\gamma\right]P_{\rm t,F} - b_{\rm F}\right)^\frac{1}{\beta}}\right];\,
    c_{2i}(\gamma)=\frac{1}{\displaystyle W_{\rm FC}\log_2\left( 1 + {hP_{\rm t,F}}\left[\textstyle\sum_{j=0}^{i-1}\gamma^{(j)}+\gamma\right] \right)}
+ \frac{c_Ca_C^{(1/\beta)} }{\displaystyle\left(P_{\rm t,C} - b_{\rm C}\right)^\frac{1}{\beta}},;\\
 & b_{2i}(\gamma)=\frac{a_{2i}(\gamma)}{\sum_{j=0}^{i}m^{(j)}{-}\sum_{j=0}^{i-1}k^{(j)}};\,
% d_{2i}(\gamma)=\frac{\sum_{j=0}^{i-1}k^{(j)}}{\displaystyle W_{\rm FC}\log_2\left( 1 + {hP_{\rm t,F}}\left[\textstyle\sum_{j=0}^{i-1}\gamma^{(j)}+\gamma\right] \right)} 
% + \frac{c_Ca_C^{(1/\beta)} \ \sum_{j=0}^{i-1}k^{(j)}}{\displaystyle\left(P_{\rm t,C} - b_{\rm C}\right)^\frac{1}{\beta}}. 
d_{2i}(\gamma)=c_{2i}(\gamma)\sum_{j=0}^{i-1}k^{(j)};\,
% \frac{\sum_{j=0}^{i-1}k^{(j)}}{\displaystyle W_{\rm FC}\log_2\left( 1 + {hP_{\rm t,F}}\left[\textstyle\sum_{j=0}^{i-1}\gamma^{(j)}+\gamma\right] \right)} 
% + \frac{c_Ca_C^{(1/\beta)} \ \sum_{j=0}^{i-1}k^{(j)}}{\displaystyle\left(P_{\rm t,C} - b_{\rm C}\right)^\frac{1}{\beta}}.
L_{2i}= \frac{\sum_{j=0}^{i}m^{(j)}}{\displaystyle W_{\rm IF}\log_2\left( 1 + {gP_{\rm t,I}}\left[\textstyle\sum_{j=0}^{i}\alpha^{(j)}\right] \right)}.
\label{eq:2i_para}
\end{align}}
\hrule
\end{figure*}
Steps for computing the solution $(s^{(i)},k^{(i)},\gamma^{(i)})$ for \eqref{eq:it_i_stage_2} are similar to those for computing $(t^{(i)},m^{(i)},\alpha^{(i)})$ in \eqref{eq:it_i_stage_1}. In particular, $k^{(i)}(\gamma) = \frac{a_{2i}(\gamma)-d_{2i}(\gamma)}{b_{2i}(\gamma)+c_{2i}(\gamma)}$ and $s^{(i)}(\gamma) = \frac{c_{2i}(\gamma)a_{2i}(\gamma) + d_{2i}(\gamma)b_{2i}(\gamma) }{b_{2i}(\gamma)+c_{2i}(\gamma)} + L_{2i}$, 
% \begin{equation}\label{eq:optimal_k_in_it_i_stage_1}
% k^{(i)}(\gamma) = \frac{a_{2i}(\gamma)-d_{2i}(\gamma)}{b_{2i}(\gamma)+c_{2i}(\gamma)} \ \mbox{and}
% \end{equation}
% \begin{equation}\label{eq:optimal_s_in_it_i_stage_1}
% s^{(i)}(\gamma) = \frac{c_{2i}(\gamma)a_{2i}(\gamma) + d_{2i}(\gamma)b_{2i}(\gamma) }{b_{2i}(\gamma)+c_{2i}(\gamma)} + L_{2i},
% \end{equation}
which are used to determine $(s^{(i)},k^{(i)},\gamma^{(i)})$. The optimal point $\gamma^{(i)}$ of \eqref{eq:it_i_stage_2} is given by
\begin{equation}\label{eq:gamma_it_i_stage_1}
\gamma^{(i)} = \argmin{ 0 \leq   \gamma  \leq  \gamma_{\textrm{max}}-\sum_{j=0}^{i-1}\gamma^{(j)}} \ s^{(i)}(\gamma),
\end{equation}
and $k^{(i)}$, $s^{(i)}$ are computed by evaluating their expressions at~$\gamma^{(i)}$.

\subsection{Sequential Latency Minimization (SLM) Algorithm}
In this section, based on the results in \S~\ref{subsec:Basis-for-the-Two-Stage-Optimization-Procedure}, we outline the sequential latency minimization (SLM) algorithm followed by its convergence properties.
\setlength{\textfloatsep}{10pt}
\begin{algorithm}[t]
\caption{SLM Algorithm}\label{Alg: SLMA}  
\begin{algorithmic}[1]
\State {\bf Initialization:} 
Set $i=1$, $(m^{(i-1)},\alpha^{(i-1)})=(0,0)$ and $(k^{(i-1)},\gamma^{(i-1)})=(0,0)$. Let $\epsilon>0$ be an accuracy~level.
\State Solve problem~\eqref{eq:it_i_stage_1} to yield $t^{(i)},m^{(i)}$ and $\alpha^{(i)}$.
\State Solve problem~\eqref{eq:it_i_stage_2} to yield $s^{(i)},k^{(i)}$ and $\gamma^{(i)}$.
\State If $|t^{(i)}-s^{(i)}|\leq\epsilon$, go to step $5$. Otherwise, set $i=i+1$ and go to step 2. 
\State {\bf Output:} Let $t^\star{=}t^{(i)}$, $m^\star=\sum_{j=0}^{i}m^{(j)}$,  $k^\star=\sum_{j=0}^{i}k^{(j)}$,  $\alpha^\star=\sum_{j=0}^{i}\alpha^{(j)}$, and  $\gamma^\star=\sum_{j=0}^{i}\gamma^{(j)}$ and~\texttt{STOP}. 
\end{algorithmic} 
\end{algorithm} %\vspace{-1em}

%\noindent\rule{1\columnwidth}{0.3mm}\vspace{-.4em}
% \\
% \emph{Algorithm}: \ \textsc{\small{Alternating Optimization Method to %Problem~ 
% \eqref{eq:equ-problem}}}\vspace{-1em}\\
%  \noindent\rule{1\columnwidth}{0.2 mm} 
% \begin{enumerate}
% 	\item Initialization: Set $i=1$, $(m^{(i-1)},\alpha^{(i-1)})=(0,0)$, $(k^{(i-1)},\gamma^{(i-1)})=(0,0)$, $A^{(i-1)}=0$, and $\Gamma^{(i-1)}=0$. Let $\epsilon>0$ be an accuracy~level.
% 	\item Solve problem~\eqref{eq:it_i_stage_1} to yield $t^{(i)},m^{(i)}$, and $\alpha^{(i)}$. %Let $A^{(i)}=A^{(i-1)}+\alpha^{(i)}$.
  %	\item If $|A^{(i)}-A^{(i-1)}|\leq\epsilon$, set $k^{(i)}=k^{(i-1)}$, $\gamma^{(i)}=\gamma^{(i-1)}$, and go to step~$7$.
%  	\item Solve problem~\eqref{eq:it_i_stage_2} to yield $s^{(i)},k^{(i)}$, and $\gamma^{(i)}$. %Let $\Gamma^{(i)}=\Gamma^{(i-1)}+\gamma^{(i)}$. 
%   	\item If $|t^{(i)}-s^{(i)}|\leq\epsilon$, go to step~$5$. Otherwise, set $i=i+1$ and go to step~2.
% 	\item Output: Let $t^\dagger{=}t^{(i)}$, $m^\dagger=\sum_{j=0}^{i}m^{(j)}$,  $k^\dagger=\sum_{j=0}^{i}k^{(j)}$,  $\alpha^\dagger=\sum_{j=0}^{i}\alpha^{(j)}$, and  $\gamma^\dagger=\sum_{j=0}^{i}\gamma^{(j)}$ and~\texttt{STOP}.
% \end{enumerate}
% \vspace{-.5em}
% \rule{1\columnwidth}{0.3mm}\vspace{-0mm}
 
The SLM algorithm is summarized in Algorithm~\ref{Alg: SLMA}. Step $1$ initializes the SLM. Steps $2$ and $3$ are the stage $1$ and stage~$2$ optimization problems, respectively. The stopping criterion is checked at step $4$. 
%
%By using the monotonicity properties of the sequences $\{t^{(i)}\}$ and $\{s^{(i)}\}$, together with some structural properties of the stage~$1$ and stage~$2$ optimizations, it can be shown that the sequence $\lim\limits_{i\rightarrow \infty }t^{(i)}-s^{(i)}=0$. Details are excluded due to space limitations. 
%
Finally, step $5$ computes the aggregate workload at the fog and cloud layers $m^\star$ and $k^\star$, together with the power split values at the IoT and fog layers $\alpha^\star$ and $\gamma^\star$, respectively. The associated latency is given by $t^\star$. The SLM algorithm always terminates after finitely many iterations, as shown in Theorem~\ref{Theorem:Convergence}, which is a consequence of following lemmas. % are used as basis for proving the theorem.

%\big(i.e., $m^\dagger$\big) and the aggregate $k$ at the cloud \big(i.e., $k^\dagger$\big), together with the power split at the IoT~\big(i.e., $\alpha^\dagger$\big) and that of~fog~\big(i.e., $\gamma^\dagger$\big). The associated latency is given by $t^\dagger$. 
%@@@@@@@@@@@@@@@@@@@@@@@@@@@@@@@@@@@@@@@@@@@@@@@@@@@@@@@@@@@@@@@@@@@@@@@@
%%%%% Commented %%%%%%
%@@@@@@@@@@@@@@@@@@@@@@@@@@@@@@@@@@@@@@@@@@@@@@@@@@@@@@@@@@@@@@@@@@@@@@@@
%\new{ 
\vspace{-1mm}
\begin{lemma}%[Comparison of $s^{(i)}$ and $t^{(i)}$]
\label{re:compare_s_n_t_n} 
 For any positive integer $i$, $t^{(i)}\geq s^{(i)}$.
\end{lemma}
\begin{IEEEproof}
This follows simply by noting that $s=t^{(i)}$, $k= 0$, and $\gamma=0$ is feasible for problem~\eqref{eq:it_i_stage_2}. Thus, the optimal value $s^{(i)}$ of problem~\eqref{eq:it_i_stage_2} no greater than~$t^{(i)}$.
\end{IEEEproof}
\begin{lemma}%[Strict Monotonicity of $\left\{t^{(i)}\right\}$ and $\left\{s^{(i)}\right\}$]
\label{re:monotone_s_n_t_n} 
The sequence ${t^{(i)}}$ is strictly monotonically decreasing and bounded below. Moreover, the sequence ${s^{(i)}}$ is strictly monotonically increasing and bounded above. 
\end{lemma}
\begin{IEEEproof}
Only an outline of the proof is provided. At the end of the first iteration,   $t^{(1)}\geq s^{(1)}$ according to {Lemma}~\ref{re:compare_s_n_t_n}. If $t^{(1)}= s^{(1)}$, the algorithm exits. Otherwise, $t^{(1)}> s^{(1)}$. Assuming this is the case, consider the second iteration. To solve the stage $1$ problem, the left-hand sides of the first two inequalities in \eqref{eq:it_i_stage_1} must be set equal, which leads to $t^{(1)}>t^{(2)} >s^{(1)}$.  Similarly, to solve the stage $2$ problem in \eqref{eq:it_i_stage_2}, the left-hand sides of the first two inequality constraints must be balanced, and thus  $s^{(1)}<s^{(2)}<t^{(2)}$. Therefore, at the end of the second  iteration, $t^{(2)}<t^{(1)}, s^{(2)}>s^{(1)}$. The iterations continue in this manner and the proof is concluded. 
\end{IEEEproof}
\begin{theorem}\label{Theorem:Convergence}
The SLM algorithm terminates in finite time. In particular, 
%\begin{eqnarray}
$\lim_{i \rightarrow \infty} \left( t^{(i)} - s^{(i)} \right) = 0.$  
%\end{eqnarray}
\end{theorem}
\begin{IEEEproof}
The proof is based on {Lemma} \ref{re:monotone_s_n_t_n}. %Details are excluded due to space limitations.
\end{IEEEproof}
%\com{
%{\em 
% {\bf \com{Questions:} }
% \begin{enumerate}
% \item We need to define $A^{(i-1)}$ and $\Gamma^{(i-1)}$ in Algorithm \ref{Alg: SLMA}.  
% \new{It is a typo propogated from previous versions. I removed it.}
% \end{enumerate}}
%  }

\vspace{-0mm}
%\newpage
%=========================================
\section{Numerical results}
In this section, numerical examples are provided to compare SLM algorithm and the optimum exhaustive search method. We consider a computing scenario in which the IoT, fog, and cloud layers are implemented with processors Quark X1000 400\,MHz, Xeon E7450 Dunnington $2.4$\,GHz, and Xeon Platinum 8156-Intel $3.6$\,GHz, respectively, with maximum power dissipations of $2.2$\,W, $90$\,W, and  $105$\,W, as given in various Intel CPU specifications. According to \eqref{e_timep}, we select $a_I$, $a_F$ and $a_C$ to satisfy these maximum powers for $\beta=3$ and $b_I=b_F=b_C=10^{-3}$. The signal-to-noise ratios (SNRs) of the links between the IoT and fog layers and the fog and cloud layers are defined as $\text{SNR}_{\rm IF} ={P_{\rm tI}}/{(N_0 W_{\rm IF})}$ and $\text{SNR}_{\rm FC} = {P_{\rm tF}}/{(N_0 W_{\rm FC})}$, respectively. %, where $N_0$ is the noise power spectral density in Watts/Hz. 
The wireless channel gain between the IoT and fog layers is exponentially distributed with unit mean. %$h_{\rm IF}$ is Rayleigh fading, i.e.,  $g_{IF}=|h_{\rm IF}|^2$ is an exponential random variable with $\mathbb{E}(g_{IF})=1$.
We set other parameters as $c_{I}=5$, $c_{F}=2$, $c_{C}=1$, $W_{\rm IF}=W_{\rm FC}=500$\,MHz, $\text{SNR}_{\rm FC}=32$\,dB, and $N_0=10^{-10}$\,Watts/Hz. 
We calculate the average latency over 4000 channel realizations. 
% Commonly used system parameters 
% are listed in Table~\ref{table_parameters}.
% %
% \begin{table}[h]
% \begin{center}
% \begin{tabular}{||l|c||l|c||}
% \hline
%  $\beta$ & $3$ &  $W_{\rm IF}$ & $500$\,MHz \\\hline
%  $C_{\rm I}$ & $5$ &  $W_{\rm FC}$ & $500$\,MHz\\\hline
% $C_{\rm F}$ & $2$&$N_0$&$10^{-10}$ Watts/Hz      \\\hline
% $C_{\rm C}$ & $1$& $\text{SNR}_{\rm FC}$ & $32$\,dB \\\hline
% %   &  & $b_{\rm I}$ &$P_{\rm t, I}/4$\\\hline
% %  &  & $b_{\rm F}$ & $P_{\rm t, F}/4$ \\\hline
% %  &  & $b_{\rm C}$ & $P_{\rm t, C}/4$\\\hline
% % $P_{\rm t, C}$ &$2(P_{\rm t, I} + P_{\rm t, F})$ &$\text{SNR}_{\rm FC}$& $20$ dB\\\hline
% \end{tabular}
% \caption{Table of parameters}
% \label{table_parameters}
% \vspace{-3em}
% \end{center}
% \end{table}
%
% %
% \begin{table}[h]
% \begin{center}
% \begin{tabular}{||l|c||l|c||}
% \hline
%  $\beta$ & $3$ &  $W_{\rm IF}$ & $500$\,MHz \\\hline
%  $C_{\rm I}$ & $5$ &  $W_{\rm FC}$ & $500$\,MHz\\\hline
% $C_{\rm F}$ & $2$&$N_0$&$10^{-10}$ Watts/Hz      \\\hline
% $C_{\rm C}$ & $1$& $B$ & $20000$ \\\hline
%  $a_{\rm I}$ & $0.5\times 10^{-9}$ & $b_{\rm I}$ &$P_{\rm t, I}/4$\\\hline
% $a_{\rm F}$ & $0.5\times 10^{-9}$ & $b_{\rm F}$ & $P_{\rm t, F}/4$ \\\hline
% $a_{\rm C}$ & $0.5\times 10^{-9}$ & $b_{\rm C}$ & $P_{\rm t, C}/4$\\\hline
% $P_{\rm t, C}$ &$2(P_{\rm t, I} + P_{\rm t, F})$ &$\text{SNR}_{\rm FC}$& $20$ dB\\\hline
% \end{tabular}
% \caption{Table of parameters}
% \label{table_parameters}
% \vspace{-3em}
% \end{center}
% \end{table}
% %
\begin{figure}[htb]
\centering
\vspace{-0ex}%
\includegraphics[width=0.7\columnwidth]
{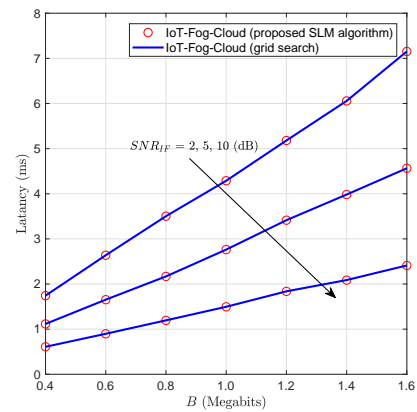}\vspace{-1em}
\caption{The average latency vs workload $B$ for different $\text{SNR}_{\rm IF}$.}
\label{fig_latency_SNR}
\vspace{-1em}
\end{figure}
\begin{figure}[htb]
\centering
\vspace{-0ex}
\includegraphics[width=0.7\columnwidth]
{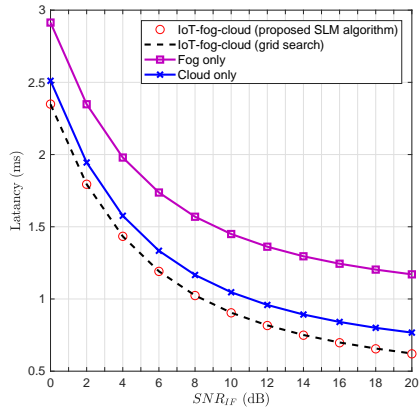}\vspace{-1em}
\caption{The average latency vs $\text{SNR}_{\rm IF}$ for different computing systems.}
\label{fig_comp}
\vspace{-1em}
\end{figure}
%
% \begin{figure}[htb]
% \begin{center}
% \includegraphics[width=0.39\textwidth]
% {fig_1_latency-v1.eps}
% \end{center}
% \caption{The average latency versus $\text{SNR}_{\rm IF}$ using the proposed method.}
% \label{fig_latency_SNR}
% \vspace{-1em}
% \end{figure}
% %
% \begin{figure}[htb]
% \begin{center}
% \includegraphics[width=0.39\textwidth]
% {fig_2_comp-v1.eps}
% \end{center}
% \caption{A comparison of the obtained latency results.}
% \label{fig_comp}
% \end{figure}
% %
% \begin{figure}[t]
% \begin{center}
% \includegraphics[width=0.39\textwidth]
% {fig_3_CI-v1.eps}
% \end{center}
% \caption{The latency versus $C_{\rm I}$ using the proposed method}
% \label{fig_CI}
% \vspace{-2em}
% \end{figure}
%

%\section{Numerical examples}
%\subsection{Performance of the Proposed SLM Algorithm}
Figure~\ref{fig_latency_SNR} shows the average latency (in milli-seconds) vs workload $B$ (in Megabits) for both the proposed SLM algorithm and the optimal grid search when $\text{SNR}_{\rm IF}=2,5,10$\,dB. %Our aim is to compare the optimal latency value of problem~\eqref{eq:equ-problem} and the latency obtained by the proposed SLM algorithm. % are compared. 
The optimum value is obtained through the exhaustive two-dimensional grid search %\footnote{At a \emph{specified} grid point $(\alpha,\gamma)$, i.e., for fixed $\alpha$ and $\gamma$, problem~\eqref{eq:equ-problem} is a linear programme with variables $m$ and $k$ and is solved optimally.} %over the valid ranges of $\alpha$ and $\gamma$ defined earlier 
with a granularity of $10^{-2}$ in each dimension.
Clearly, the results of both methods coincide, which suggests that the SLM algorithm performs very close to the optimum method. % optimally for given set of parameters.
Figure \ref{fig_latency_SNR} also indicates that the latency increases almost linearly with workload $B$. % and decreases when $\text{SNR}_{\rm IF}$ increases. 
For example, for the simulated range at $\text{SNR}_{\rm IF}=5$\,dB, latency increases from $1.1$\,ms to $4.5$\,ms, where we need $2.9$\,ms to process one Megabits of data.  
Further, to achieve 2\,ms latency, we can process approximately 0.45, 0.75 and 1.35 Megabits when $\text{SNR}_{\rm IF}=2,5,10$\,dB, respectively.

% Figure shows that the average latency decreases when $\text{SNR}_{\rm IF}$  increases, as expected. A similar trend of the average latency can be observed when $\text{SNR}_{\rm FC}$ increases for fixed $\text{SNR}_{\rm IF}$. 
% \vspace{-0.3em}
% \subsection{Comparison of Different Computing Architectures}
Figure~\ref{fig_comp} depicts the average latency (in milli-seconds) vs $\text{SNR}_{\rm IF}$ when workload $B=1$\,Megabits. It  compares three different architectural choices: i) IoT-fog-cloud; ii) fog-only; and iii) cloud-only. The average latency decreases when $\text{SNR}_{\rm IF}$  increases, as expected. Results shows that the IoT-fog-cloud computing architecture always outperforms others.  For example, to yield a $1$\,ms latency, the IoT-fog-cloud computing system requires $\text{SNR}_{\rm IF}=8$\,dB, whereas the cloud-only computing system needs $\text{SNR}_{\rm IF}=11$\,dB. The fog-only computing system cannot yield a $1$\,ms latency even when $\text{SNR}_{\rm IF}=20$\,dB. The IoT-fog-cloud computing architecture always yields a decrease in the latencies, irrespective of $\text{SNR}_{\rm IF}$. For example, at $\text{SNR}_{\rm IF}=16$\,dB, the increase in latencies of the fog-only and cloud-only computing systems, compared to the IoT-fog-cloud computing system is $79$\% and $21$\%, respectively. %Results suggests that the joint utilization of computing resources at IoT, fog, and cloud layers is always better than single-point computing.%\footnote{Depending on parameter selection, the performance grade %of fog-only and cloud-only computing 
\section{Conclusion}\label{s_con}
%==============================================================================

The power and workload allocation problem to minimize data processing latency for a three-layer IoT-fog-cloud computing systems was investigated. The resulting problem is non-convex. To devise an efficient solution method, a constraint relaxation was considered yielding, under reasonable grounds, \emph{a very good} approximation to the original problem formulation. A sequential latency minimization (SLM) algorithm based on alternating optimization was proposed to handle the relaxed problem. Convergence of the SLM algorithm was established. Numerical results suggested that the performance of SLM algorithm was almost identical to that of the optimum exhaustive search method for the relaxed problem. Finally, we evaluated numerically the gains of the three-layer IoT-fog-cloud computing over fog-only and cloud-only computing, in terms of data processing latencies. Results suggest that the three-layer computing is more potent, for yielding better latencies, than fog-only or cloud-only computing systems.

\vspace{-0.25em}
%=====================================

%==============================================================================
%\appendix \label{a_proof1}
%==============================================================================

\end{document}